2013 International Conference on Future Trends in Computing and Communication Technologies2013 International Conference on Future Trends in Computing and Communication Technologies

# A Novel Application Licensing Framework for Mobile Cloud Environment

Atta ur Rehman Khan, Mazliza Othman, Abdul Nasir Khan
Faculty of Science and Information Technology, University of Malaya, Kuala Lumpur, Malaysia

Email: attaurrehman@siswa.um.edu.my

*Abstract*— Mobile cloud computing is a new technology that enhances smartphone applications capabilities in terms of performance, energy efficiency, and execution support. These features are achieved via computation offloading technique that is supported by specialized mobile cloud application development models. However, the cloud-enabled applications are prone to application piracy issue for which the traditional licensing frameworks are of no use. Therefore, a new licensing framework is required to control application piracy in mobile cloud environment. This paper presents a preliminary design of a novel application licensing framework for mobile cloud environment that restricts execution of applications on unauthenticated smartphones and cloud resources.

*Keywords– application piracy, mobile cloud computing, mobile cloud applications piracy, mobile cloud application models, mobile cloud data privacy.*

## I. INTRODUCTION

The term *cloud computing* was introduced in early 2008. Cloud computing provides flexible and virtually unlimited resources/services via the Internet. Since its foundation, researchers from all over the world are investigating potential applications for this technology. At the same time, smartphones have also achieved high popularity due to support for applications from various domains, such as entertainment, e-commerce, education, healthcare, and safety. These applications incur an ever increasing computational and energy demands on the smartphones [1].

To meet these demands, researchers envision the usage of cloud computing for mobile devices, which gives birth to a new domain called mobile cloud computing (MCC). MCC can be defined as *"an integration of cloud computing technology with mobile devices to make the mobile devices resource-full in terms of computational power, memory, storage, energy, and context-awareness"* [1]. Figure 1 shows a high level architecture of mobile cloud.

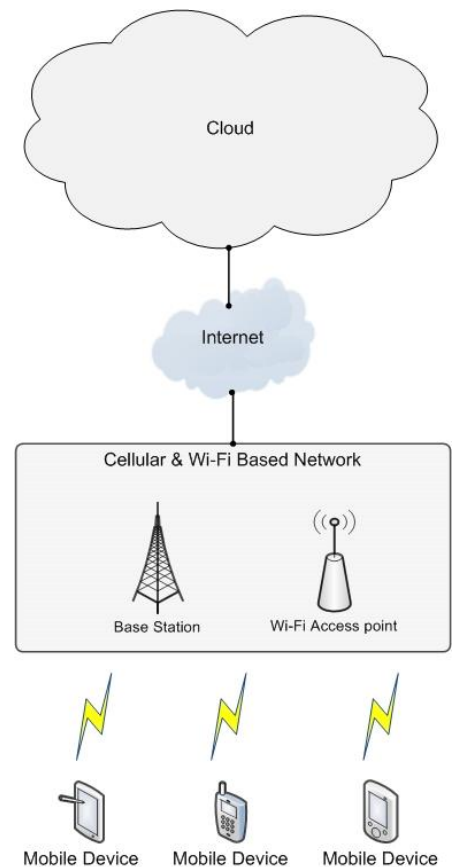

Figure 1: A high level architecture of mobile cloud

MCC is an emerging technology that uses cloud resources in mobile environment to virtually enhance smartphone's resources and overcome the long lasting challenges, such as limited energy, computational power, and execution support for recourse intensive applications [2]. Studies show that MCC has succeeded in catering to many smartphone constraints by offloading the resource intensive computational tasks to the cloud [1, 3, 4]. However, to enable computation offloading from the smartphones, the applications have to be developed using specialized mobile cloud application development models.

These models facilitate in the development of smartphone applications, which can leverage performance, energy efficiency, and execution support by utilizing cloud recourses. To do so, cloud-enabled applications send computational requests to the cloud, where the computations are performed, and the results/data are synced back to the smartphone via an Internet connection.

Mobile cloud application development models face many challenges and are affected by multiple issues, which are discussed by Khan *et al.* in [1]. In this paper, we highlight the application piracy issue in application models and propose a novel application licensing framework for mobile cloud environment named Pirax. Piracy control of cloud-enabled applications is important because it not only causes profit loss to the application providers, but also incurs additional cost in terms of cloud service usage.

The rest of the paper is organized as follows. Section II highlights the related work. Section III presents challenges and assumptions. Section IV discusses the proposed framework, and Section V concludes the paper.

## II. RELATED WORK

The application models offload computations to the cloud by means of a process, component, application, or a smartphone clone. Based on the offloading techniques, the application models [5-10] can be classified into two main types, namely augmented execution and elastic application models.

### A. Augmented execution models

In these models, a smartphone clone image is generated containing installed applications and data, and is transferred to the cloud, where it resides and executes on a virtual machine (VM) [5-7, 9]. The clone facilitates smartphone applications in executing the computational requests and performs periodic/runtime synchronization with the smartphone for consistent execution.

### B. Elastic application models

The applications developed with this type of models are divided into sub-partitions, called components, which can execute independently on a smartphone or cloud. The components that require user interaction or local resources (GPS, camera, sensors) are left on the smartphone, whereas resource intensive components are offloaded to the cloud [8].

Figure 2 presents the execution of applications based on elastic application and augmented execution based models.

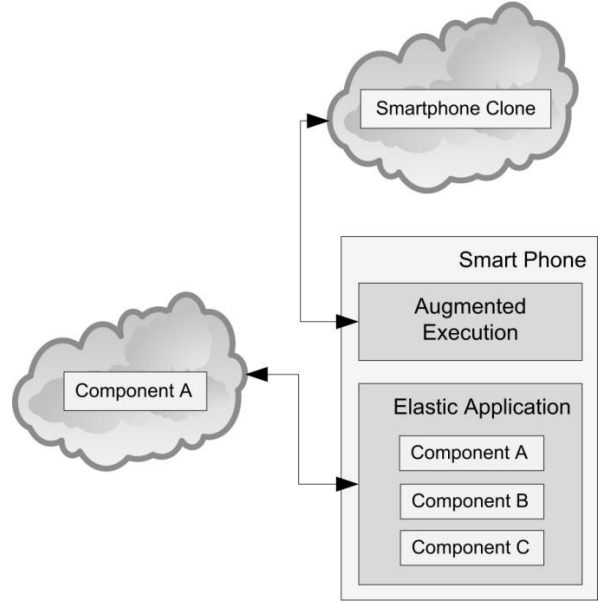

Figure 2: Execution of applications based on elastic application and augmented execution based models.

The augmented execution based models introduce three main challenges, namely synchronization of smartphone and clone, privacy of data, and piracy of applications, which are explained as follows [1]:

(i) The synchronization of smartphone and clone is important for consistent execution of applications in the cloud. Therefore, whenever computation offloading is required, it is ensured that the smartphone and its clone are in synchronized state. However, depending on the type of synchronization (periodic, runtime), this activity may incur high computational delays and bandwidth consumption.

(ii) The smartphone clone may contain confidential user data, which raises data privacy concerns.

(iii) The clone may contain licensed applications, which may give rise to application piracy issue.

### C. Threat model

In mobile cloud environment, the computations are performed on unencrypted data. Therefore, if a hacker obtains a copy of the smartphone clone, then the user's confidential data can be extracted. Moreover, the clone can be installed on a smartphone of the same model to illegally access the installed applications. Furthermore, the applications can be extracted along with any dependencies for illegal distribution, causing application piracy.

Unfortunately, the existing mobile cloud application models assume that cloud is secure, and hence, no efforts are made to secure the clone or its applications. In addition, there is no mechanism to ensure privacy of data by restricting the execution of applications on unauthenticated smartphones and cloud resources.

Considering these threats, it is vital to control application piracy in mobile cloud environment by revoking execution access on unauthenticated devices/VMs. The execution control of applications will not only facilitate in controlling application piracy, but may also increase the privacy of data, where the application's data is stored in a custom format and is only accessible via the application.

## III. CHALLENGES & ASSUMPTIONS

The primary objective of Pirax is to control application piracy in mobile cloud environment. The challenges and assumptions regarding the design and development of Pirax are discussed as follows:

### A. Challenges

In traditional licensing frameworks [11], the application license is usually assigned on the basis of smartphone's unique hardware entity, namely the International Mobile Equipment Identity (IMEI) [12]. These frameworks work quite well, because the applications always execute on a smartphone and there is no requirement for application mobility. On the other hand, cloud-enabled applications (for mobile cloud computing) can execute on the smartphone or migrate to the cloud to utilize cloud resources. Consequently, cloud-enable applications require a licensing framework that can lock applications to a unique smartphone and cloud entity.

To the best of our knowledge, there exists not a single licensing framework for mobile cloud environment. Therefore, we need to identify unique entities in the cloud environment that can facilitate in binding the application license. After investigating various cloud entities, we found that every VM instance in the cloud has a Universally Unique IDentifier (UUID) that tracks VM instances [13]. Therefore, with minor configuration of the virtualization software [14], [15], it can be assured that a user receives the same UUID for the assigned VM instance, every time.

### B. Assumptions

Pirax makes the following assumptions:

- A smartphone user is assigned a cloud-VM with unique UUID, which do not change on VM restart, migration, and upgrade.

- The Smartphone Activation Request Serial (SARS), Smartphone Activation Serial (SAS), Cloud Activation Request Serial (CARS), and Cloud Activation Serial (CAS) (discussed in the next section) are communicated between the smartphone/VM and application provider in encrypted form. Therefore, information of IMEI, UUID, SAS, or CAS cannot be compromised.

- The encoding modules used on the smartphone/VM for SARS/CARS generation, and application provider for SAS/CAS generation are secure and cannot for replicated.

## IV. PROPOSED MODEL - PIRAX

In order to bind an application to a smartphone and VM instance in the cloud, Pirax generates two licenses, i.e., one for smartphone and one for cloud.

### A. Smartphone license

When the smartphone application executes for the first time, Pirax fetches the IMEI of the smartphone and generates SARS. Pirax sends the SARS to the application provider, where it generates SAS based on the smartphone's IMEI. On receiving the SAS from application provider, the smartphone application compares it with the IMEI information. If it is a match, then Pirax activates the application and stores SAS for validating application executions. SARS and SAS are generated based on two different encoding schemes, due to which, the IMEI information cannot be extracted from the SAS. Moreover, even if the IMEI information is known, new SAS cannot be generated for a different smartphone.

To validate the execution of an application, Pirax retrieves the IMEI information and compares it with the SAS. If the information contained in the SAS and smartphone IMEI matches, then the application is allowed to execute. Otherwise, Pirax considers the smartphone as an unauthenticated device and restricts the application from execution. Figures 3 and 4 present the license generation process and Pirax architecture for the smartphone.

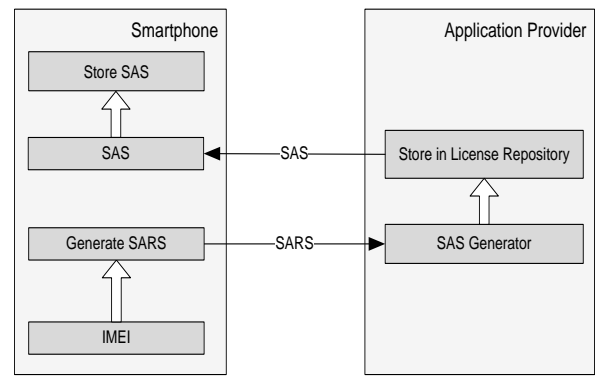

Figure 3: Smartphone license generation process

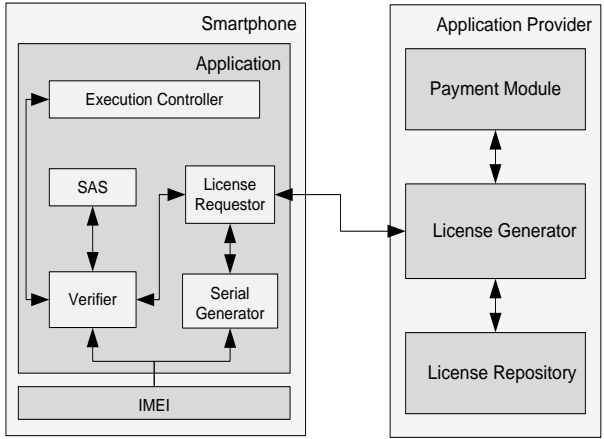

Figure 4: Pirax architecture for the smartphone

*B. Cloud license*

For cloud license generation, Pirax generates CARS based on the UUID of the VM and SAS of the smartphone. Pirax uses SAS in the generation of CARS, because it tracks the user and the respective license type (smartphone only, smartphone and cloud). Similar to the smartphone case, the application provider generates a license and sends it to the smartphone clone in the cloud. Pirax compares the information of CAS with UUID and SAS (on clone), and stores CAS to validate application executions on the VM.

When the application executes in the cloud, Pirax compares CAS with the UUID and SAS for license validation. If the information contained in the CAS matches the UUID and SAS, then the application is allowed to execute in the cloud. Otherwise, it considers the VM as an unauthenticated instance and restricts the application from execution. Figures 5 and 6 present license generation process and Pirax architecture for the cloud.

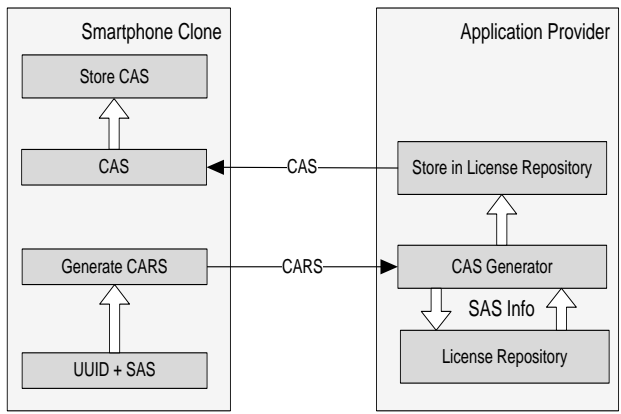

Figure 5: Smartphone license generation process

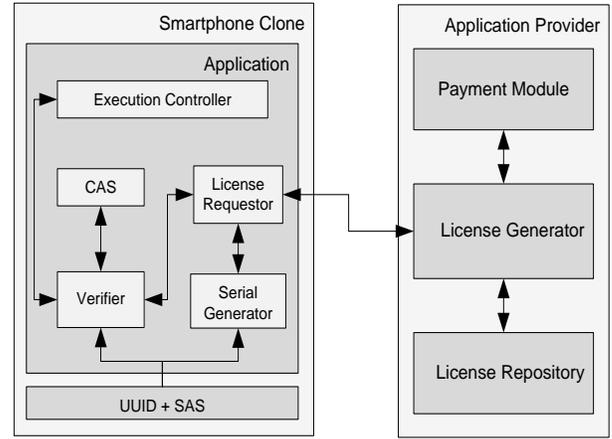

Figure 6: Pirax architecture for the cloud

As the copy of SAS and CAS are stored by the application provider, Pirax allows re-installation and activation of application on the same smartphone and cloud instance.

## V. CONCLUSION

In this paper, we propose a novel licensing framework for application piracy control in mobile cloud environment. Pirax facilitates to control application piracy on the smartphone as well as in the cloud environment. Moreover, it provides an additional benefit of data privacy by restricting the applications from executing on unauthenticated smartphone and cloud resource. Pirax is easy to integrate into existing mobile cloud application models. For future work, we intend to validate the model using petri-nets [16] and implement it for the Android platform [17].